\documentclass[twocolumn,twoside,slac_two]{revtex4}
\usepackage{graphicx}
\usepackage{fancyhdr}
\begin{document}

\title{The VERITAS Extragalactic Science Program}

%

\author{N. Galante}
\affiliation{Harvard-Smithsonian Center for Astrophysics}
\author{for the VERITAS Collaboration}
\affiliation{http://veritas.sao.arizona.edu}

\begin{abstract}
VERITAS is an array of four 12-m diameter imaging atmospheric-Cherenkov telescopes located in southern Arizona. Its aim is to study the very high energy (VHE: $E>100$~GeV) $\gamma$-ray emission from astrophysical objects. The study of Active Galactic Nuclei (AGN) is intensively pursued through the VERITAS blazar key science project, but also through the large MWL observational campaigns on radio galaxies. The successful VERITAS AGN research program has provided insights to the jet inner structures and started a more detailed classification of blazars. Moreover, the synergy between Fermi and VERITAS on blazar observations resulted in important constraints on the extragalactic background light (EBL) through $\gamma$-ray observations, and on the cataloguing of the several AGN sub-classes . VERITAS discovered also the first extragalactic non-AGN $\gamma$-ray source. The discovery of gamma-ray emission from the starburst galaxy M~82 by VERITAS and Fermi provides important clues on possible mechanisms for accelerating cosmic rays.
\end{abstract}

\maketitle

\thispagestyle{fancy}


\section{Introduction}

The largest population of VHE-detected $\gamma$-ray sources
are blazars, a sub-category of AGNs in which the ultra-relativistic jet, produced by the
accretion of matter around a super-massive black hole, is aligned within a few degrees to the
observer's line of sight~\cite{UrryPadovani}. The most commonly accepted model
to account for the $\gamma$-ray emission from the jet of AGNs is inverse-Compton
scattering of the synchrotron photons produced by shock-accelerated electrons and positrons
within the jet itself~\cite{Jones1974}. In these aligned sources relativistic beaming substantially
boosts the apparent flux.
Non-blazar AGNs are typically oriented at 
larger angles from the observer's line of sight, becoming much more challenging. However,
the misalignment enables imaging of the jet's structure, crucial to identifying the emission regions
and probing models of the acceleration mechanism. Given the typical angular resolution
of the order of several arcminutes in $\gamma$-ray instruments, jet substructures are not resolved
in the $\gamma$-ray energy band, but  they are in other wavelengths.
Correlation studies through coordinated
multi-wavelength (MWL) observational campaigns on radio galaxies are a viable strategy
to investigate the physical processes at work in the substructures of the jet. A dedicated contribution about radio
galaxies is presented in these proceedings.

The advantage of observing radio galaxies is that it is possible to study also the rich environment
in which they are typically located. It has been seen that radio galaxies are preferentially
located in cluster of galaxies~\cite{Prestage1988}. Their powerful jets energize the intra-cluster medium through the 
termination shocks accompanied by particle acceleration and magnetic field amplification. 
Large scale AGN jets and cluster of galaxies are believed to be potential accelerator for cosmic 
rays~\cite{Dermer2009}, therefore the modeling of the dynamics of both populations is of particular 
interest for the cosmic-ray community.

Beside AGN-related environments,
starburst galaxies are also good candidates as ultra-high energy cosmic rays accelerators. The active
regions of starburst galaxies have a star formation rate about 10 times larger than the rate
in normal galaxies of similar mass, with a consequent higher rate of novae and supernovae.
The cosmic rays produced in the formation, life, and death of their massive stars are
expected to eventually produce diffuse gamma-ray emission via their interactions with interstellar 
gas and radiation.

Finally, globular clusters are the closest extragalactic structures whose physics is interesting
to the $\gamma$-ray community. They can host hundreds of millisecond pulsars
which can accelerate leptons at the shock waves originating in collisions of the pulsar winds 
and/or inside the pulsar magnetospheres. Energetic leptons diffuse gradually through the 
globular cluster. Comptonization of stellar and microwave background radiation is therefore
expected to be responsible of $\gamma$-ray emission.

The indirect search for dark matter (DM) candidates,
 is also part of the VERITAS extragalactic non-blazar program. A dedicated contribution
on the VERITAS DM program is presented in a separate proceeding.
Highlights on the research topics and results of the VERITAS extragalactic non-blazar science
program are here presented.



\section{The Extragalactic Science Program}

\subsection{Blazars}

Since the beginning of its operations in October 2007, VERITAS observation of blazars averaged
$\sim410$~hr per year, resulting in the detection of 20 blazars (15 HBL and all 5 known IBL),
including 10 discoveries.

\begin{figure*}[!t]
  \vspace{5mm}
  \centering
  \includegraphics[width=6.0in]{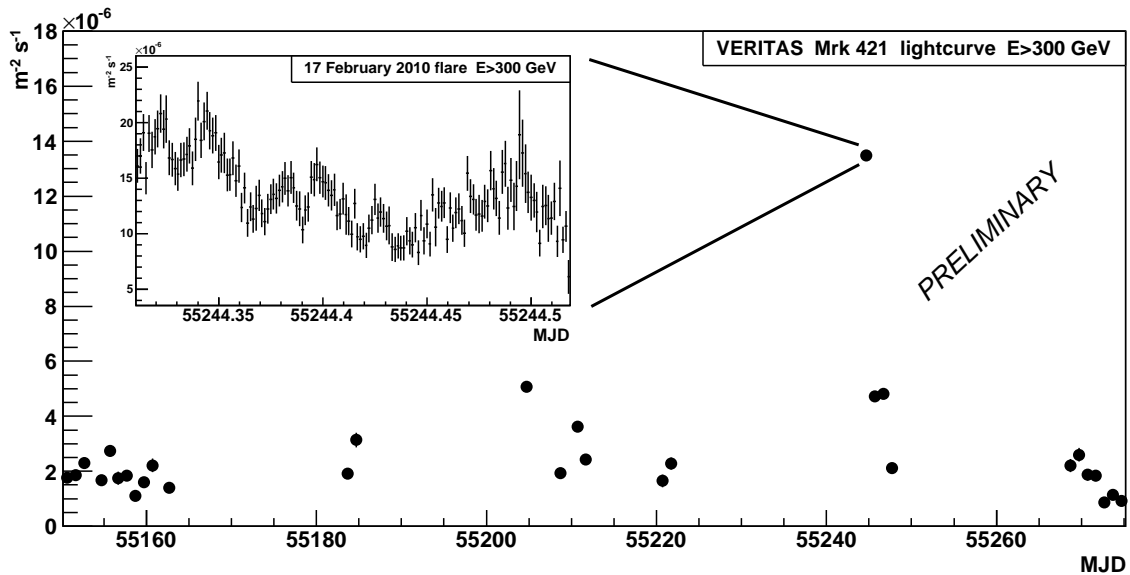}
  \caption{Nigthly lightcurve of Markarian~421 during the VERITAS 2009-2010 monitoring campaign. The source is detected at an average flux level of approximatively 2 Crab units over the entire season. On February 17, 2010 the source is observed in flaring state at a flux level of $\sim8$ Crab units. A zoom of the intra-night 2-minute bins lightcurve of the
flaring event is shown. During the flaring event the source and intra-night variability is seen.}
  \label{fig_Mrk421}
 \end{figure*}

\begin{figure}[!t]
  \vspace{5mm}
  \centering
  \includegraphics[width=3.0in]{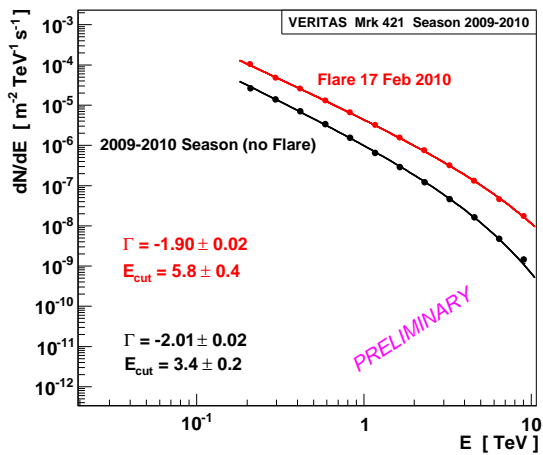}
  \caption{VERITAS preliminary spectral analysis of Markarian~421 for two different flux levels: the flare on February 17, 2010 (red dots) and the rest of the 2009-2010 season (black dots).}
  \label{fig_Mrk421spec}
 \end{figure}

Mrk~421 is the longest-known VHE blazar, and generally
has the brightest VHE flux. It is easily the best-studied
HBL at VHE, and VERITAS has acquired nearly 80~hr on
this blazar since 2007, largely during flaring states identified with the Whipple 10-m telescope. A total of 47~hrs of
VERITAS and 96~hrs of Whipple 10-m data taken between
2006 and 2008 are presented in~\cite{Acciari2011}. 
VERITAS monitoring of the VHE flux from Mrk~421 in 2009-10 reveals
an elevated state during the entire season. An extreme flare was observed for nearly 5~hr live time on
February~17, 2010, during which the VHE flux averaged
$\sim8$~Crab and showed variations on timescales of approximately 5-10 minutes [8]. 
Figure~\ref{fig_Mrk421} shows the seasonal light curve with a zoom on the night of the flare.
Variability on the time scale of few minutes is visible. Figure~\ref{fig_Mrk421spec}
shows a preliminary spectral analysis for two different flux levels: the flare happened
on February 17, 2010, and the rest of the season 2009-2010. Clear evolution
of the spectral parameters can be seen.

H 1426+428 was first detected during an outburst in 2001~\cite{Horan2002}. This HBL was observed by VERITAS for 22~hr
quality-selected live time between 2007 and 2011. A weak
excess, $5.2\sigma$, is observed in these data, marking the first
time H~1426+428 is detected since 2002. The observed
flux is <2\% Crab, well below the value (13\% Crab) reported during its VHE discovery, and also below any other
published VHE flux or limit from this source.

PG 1553+113 is a hard-spectrum ($\Gamma_\mathrm{LAT} \sim -1.66$) Fermi-LAT blazar~\cite{Abdo2010}. 
It is likely the most distant HBL detected
at VHE (see $z > 0.43$ from [3]). It was observed by VERITAS for 60 h of quality-selected live time between May
2010 and May 2011. These data result in the most signiÞcant VHE detection ($39\sigma$) 
of this HBL. The time-averaged
VHE ßux is 10\% Crab above 200~GeV, higher than the
archival VHE measurements, and the photon spectrum is
well described between 175~GeV and 500~GeV by a powerlaw function with photon index 
$\Gamma = -4.41\pm0.14$. The VHE
spectrum can be used to set an upper limit on the redshift
of $z < 0.5$. 

1ES 0229+200 is one of the hardest-spectrum VHE blazars
known ($\Gamma_\mathrm{HESS} = - 2.5$; [18]). It was observed by VERITAS as part of an intense MWL observation campaign for
46~h live time from 2009-11. A strong signal is detected
($\sim600$ $\gamma$-rays, $9.0\sigma$) in these observations corresponding
to an average VHE flux of $\sim2$\% Crab above 300~GeV. The
VERITAS flux is variable on a timescale of months, and the
preliminary VHE spectrum measured between 220~GeV
and 15~TeV has photon index $\Gamma = -2.44 \pm 0.11$. The
results of the MWL campaign are in preparation. It is interesting to note that 1ES~0229+200 is the only VERITAS-detected blazar not included in the 1FGL catalog~\cite{Abdo2010}.

1ES 0414+009 is the most distant VHE
HBL with a well-measured redshift ($z = 0.287$). It was
observed by VERITAS for 55~h of quality-selected live
time from January 2008 to February 2011. An excess
of VHE $\gamma$-rays is detected ($\sim7\sigma$) from this Fermi-LAT
source ($\Gamma_\mathrm{LAT} = - 1.94$;~\cite{Abdo2010}). The observed VERITAS
spectrum between 230~GeV and 1.8~TeV is relatively
hard ($\Gamma = 3.4 \pm 0.5$) considering EBL-related effects, and
consistent with that observed during the HESS discovery
[19]. The observed VERITAS flux is somewhat higher
(1.6\% Crab) than measured by HESS (0.6\% Crab above
200~GeV), although the large datasets used by both experiments are not simultaneous. Results from a contemporaneous MWL observation campaign are in preparation.

B2 1215+30, an IBL discovered at VHE during a flare in
January 2011~\cite{Mariotti2011}, was observed
for 55~h of qualityselected live time between December 2008 and April 2011.
The measured excess of $\sim240$ $\gamma$-rays ($6.3\sigma$) corresponds to
a VHE flux of $\sim1$\% Crab. There is a weak indication that
the flux observed by VERITAS in 2011 may be higher than
seen from 2008-10. The VERITAS flux is consistent with
that ($2 \pm 1$\% Crab) reported during the MAGIC discovery

\subsection{Radio Galaxies}

Radio galaxies observed by VERITAS include M~87, 3C~111 and NGC~1275.
A dedicated contribution on M~87 is presented in these proceedings. 
A preliminary analysis of 11~hr of quality-selected data of 3C~111 results in a flux upper limit
of $\sim3$\% Crab flux above 200~GeV.
NGC~1275 is an unusual early-type galaxy located in the center
of the Perseus cluster. Its radio emission is core dominated, but emission lines are also seen, making
it difficult to classify it according to the Faranhoff \& Riley (FR) classification~\cite{FR}. In Fall 2008 the
\emph{Fermi} $\gamma$-ray space telescope reported the detection of $\gamma$-ray
emission from a position consistent with the core of NGC~1275. VERITAS observed the core region of 
NGC~1275 for about 11~hr between 2009 January 15 and February 26, resulting in 7.8~hr of quality-selected
live time. No $\gamma$-ray emission is detected above the analysis energy threshold of $\sim$190~GeV,
resulting in a flux upper limit incompatible with the extrapolation of the \emph{Fermi-LAT} spectrum.
Under the assumption of a SSC emission mechanism, the VERITAS result suggests the presence of a cutoff
in the sub-VHE energy range~\cite{Galante}. The detection in Summer 2010 of VHE $\gamma$-ray emission 
by MAGIC~\cite{MAGIC1275} has eventually included NGC~1275 among the few interesting radio galaxies
for future VHE investigation.

\subsection{Clusters of Galaxies}

Observation of clusters of galaxies is done unavoidly during the observation of many radio galaxies. 
This is the case for NGC 1275 and M 87 where the Perseus and Virgo clusters respectively are observed during the 
radio galaxy observation. However, up to now a dedicated study of the clusters themselves
 has been done only on the 
Coma cluster. The Coma cluster is a nearby cluster of galaxies which is well studied at all 
wavelengths~\cite{Neumann2003}. 
It is at a distance of 100~Mpc ($z=0.023$) and has a mass of $2 \times 10^{15}\; M_{\odot}$. 
Its X-ray and radio features suggest the presence of accelerated electrons in the intergalactic medium 
emitting non-thermal radiation. Beside relativistic electrons, there may also be a population of 
highly energetic protons. Both high energy electrons and protons are known to be able to produce VHE photons.
A total of 19 hr of data have been recorded between March and May 2008.
No evidence for point-source emission was observed within the field of view and a preliminary upper limit of 
$\sim$3\% of the Crab flux is given for a moderately extended region centered on the core~\cite{Perkins}.

\subsection{Starburst Galaxies}

M~82 a prototype small starburst galaxy, located approximately 3.7~Mpc  from Earth, 
in the direction of the Ursa Major constellation. M~82 is gravitationally interacting 
with its nearby companion M~81. This interaction has deformed M~82 in such a way that an active starburst 
region in its center with a diameter of $\sim$1000 light years has been developed~\cite{Yun1994,Volk1996}.
Throughout this compact region stars are being formed at a rate approximately 10 times faster 
than in entire ``normal" galaxies like the Milky Way. Hence the supernovae rate is 0.1 to 0.3 per 
year~\cite{Kronberg1985,Fenech2008}.
The high star formation rate in M 82 implies the presence of numerous shock waves in 
supernova remnants and around massive young stars. Similar shock waves are known to 
accelerate electrons to very high energies, and possibly ions too. 
The intense radio-synchrotron emission observed in the central region of M 82 suggests a very 
high cosmic-ray energy density, about two orders of magnitude higher than in the Milky Way~\cite{Rieke1980}.
Acceleration and propagation
of cosmic rays in the starburst core are thus expected to be responsible for VHE $\gamma$-ray
emission.
Theoretical predictions include significant contributions from both leptonic and hadronic particle interactions.
Cosmic-ray ions create VHE gamma rays through collisions with interstellar matter,
producing $\pi^0$ which decay into $\gamma$-rays. Alternatively, accelerated cosmic-ray electrons may
inverse-Compton scatter ambient X-ray photons up to the VHE 
range~\cite{Volk1996,Pohl1994,Persic2008,deCea2009}. 

VERITAS observed M~82 for a total of ~137 hours of quality-selected live time between 
January 2008 and April 2009 at a mean zenith angle of 39$^\circ$.
An excess of 91 gamma-ray-like events ($\sim$0.7 photons per hour) are detected for a total
4.8$\sigma$ statistical post-trials significance above 700~GeV. The observed differential 
VHE gamma-ray spectrum is best fitted using a power-law function with a photon index 
$\Gamma = 2.5 \pm 0.6_\mathrm{stat} \pm 0.2_\mathrm{sys}$.
Comparison
of the VERITAS VHE spectrum with predictions of the theoretical models supports a hadronic
scenario as the dominant process responsible for the VHE emission~\cite{Acciari2009}.

\section{Conclusions}

The VERITAS extragalactic science program is well established.
The study of the AGN physics develops through two complementary observational programs of blazars 
and non-blazar (radio galaxies) AGN. Several blazars have been detected or discovered,
and a new catalog of a sub-class of AGN, the IBL catalog, was started by VERITAS. The VHE-IBL catalog
consists entirely in VERITAS-discovered blazars. Distant blazars are also studied, resulting in significant measurements for the EBL characterization.
Non-AGN sources are also investigated. A dedicated VHE study on the Coma cluster of galaxies resulted in a flux 
upper limit on the extended region centered on the core.
The first detection of $\gamma$-ray emission from a starburst galaxy established a connection
between cosmic-ray acceleration and star formation.\\

This research is supported by grants from the US Department of Energy, the US National Science Foundation, 
and the Smithsonian Institution, by NSERC in Canada, by Science Foundation Ireland, and by STFC in the UK. 
We acknowledge the excellent work of the technical support staff at the FLWO and at the collaborating 
institutions in the construction and operation of the instrument.


\end{document}